\documentclass[%
aps,
  prfluids,
  ]{revtex4-2}

  \usepackage{graphicx}
  \usepackage{dcolumn}
  \usepackage{bm}
  \usepackage{comment}
  \usepackage{graphicx}
  \usepackage{ascmac}
  \usepackage{amsmath,amssymb}
  \usepackage{mathtools}
  \usepackage{color}
  \usepackage{xcolor}
  \usepackage{overpic}
  
      \usepackage{amsthm}
    \theoremstyle{definition}
    \newtheorem{dfn}{Definition}[section]
    \newtheorem{prop}[dfn]{Proposition}

\usepackage[whole]{bxcjkjatype}
  \begin{document}

\title{Localization and bistability of bioconvection in a doubly periodic domain}

  \author{Yoshiki Hiruta}
  \email{hiruta@kurims.kyoto-u.ac.jp}
  \author{Kenta Ishimoto}
  \email{ishimoto@kurims.kyoto-u.ac.jp}
  \affiliation{
    Research Institute for Mathematical Sciences, Kyoto University, Kyoto 606-8502, Japan}
  \date{\today}

\begin{abstract}
A suspension of swimming microorganisms often generates a large-scale convective pattern known as bioconvection. In contrast to the thermal Rayleigh-B\'enard system, recent experimental studies report an emergence of steady localized convection patterns and bistability near the onset of instability in bioconvection systems. In this study, to understand the underlying mechanisms and identify the roles of particle self-propulsion in pattern formation, we theoretically and numerically investigate a model bioconvection system in a two-dimensional periodic boundary domain. In doing so, we extend a standard bioconvection model by introducing the equilibrium density profile as an independent parameter, for which the particle self-propulsion is treated as an independent dimensional parameter. Since the large-scale vertical structure dominates in this system, we are able to simplify the model by truncating the higher vertical modes. With this truncated model, we analytically derived the neutrally stable curve and found that the particle motility stabilizes the system. We then numerically analyzed the bifurcation diagram and found the bistable structure at the onset of instability. These findings, localization and bistability, are consistent with experimental observations. We further examined the global structure of the bistable dynamical system and found that the non-trivial unstable steady solution behaves as an edge state that separates the basins of attraction. These results highlight the importance of particle self-propulsion in bioconvection, and more generally our methodology based on the dynamical systems theory is useful in understanding complex flow patterns in nature.
\end{abstract}

\maketitle
\section{Introduction\label{sec:introduction}}

A suspension of swimming microorganisms often forms convection flows at the mm-cm length scales and these self-organized patterns, known as bioconvection, have attracted researchers more than a half century \cite{childress_pattern_1975,pedley_1992}.
The bioconvection is known to driven by biased self-propelled locomotion of microscopic particles \cite{bees_2020}, %
such as phototaxis of phytoplankton, chemotaxis of bacteria and gyrotaxis of bottom-heavy swimmers.

These microorganisms are typically heaver than the surrounding fluid and when they accumulate near the top of the container, density overturning instability may occur, analogous to Rayleigh-B\'enard thermal convection. It is also known that a bottom-heavy microswimmer can form a convection pattern without an upper surface or vertical density gradient \cite{pedley_growth_1988}.

Among various experimental and numerical studies on bioconvection, Shoji et al. \cite{shoji_2014} experimentally studied a suspension of {\it Euglena gracilis} in an annular container, which confined the fluid motion almost in two dimensions, and found that the suspension of {\it E. gracilis} formed a spatially localized convection pattern under strong light from the bottom of the container. 
They also reported that the emergence of the local flow structure depends on  initial cell concentrations and thus concluded that the system is bistable in the sense that the base flow and convection flow are both stable states.

Bistability is often observed in the dissipative system, including fluid dynamics and the reaction-diffusion system
\cite{POMEAU_1986}.
A notable example is the laminar-turbulence transition in a wall-bounded flow \cite{eckhardt_2007,avila_2023}. 
In a pipe flow, although the laminar flow solution is considered to be linearly stable at an arbitrary large Reynolds number, the flow experimentally becomes turbulent at a finite Reynolds number $(\approx 2000)$.
From dynamical systems' points of view, the laminar and turbulent states are bistable solutions of the incompressible Navier-Stokes equations. In this Reynolds number regime, spatially localized turbulent flow patterns, called puffs, appear as a bistable state \cite{avila_2023}. 
More recently, Hiruta \& Toh \cite{hiruta_solitary_2015,hiruta_intermittent_2017,hiruta_subcritical_2020} studied a simple shear system without a wall boundary, known as the Kolmogorov flow, and found that a localized solution is realized accompanied by the system bistability.
The bistable structure between the conduction and convection states, together with an emergence of localized solutions, is also reported in a convection of binary fluids \cite{Watanabe_2012}.

In contrast to these systems, in the Rayleigh-B\'enard thermal convection, as the temperature difference increases,
the conduction state with a linear temperature profile becomes unstable to generate a convection state with spatially periodic roll patterns. In addition,  
the conduction and convection states cannot be bistable due to the self-adjoint property of the system equation \cite{joseph_1965,joseph_1966}. 
These differences emphasize the importance of self-propulsion in forming the localized convection patters and the bistability of the system.

Hence, in this study, by examining stability of a model bioconvection system, we aim to understand how particle self-propulsion gives rise to a localized flow pattern and system bistability. Here, we recall that the surface of the container affects the density profile of the base state that corresponds to the conduction state. In turn, the fluid-density coupling that triggers the convection pattern is inevitably masked by the equilibrium density profile. 
To resolve this complexity, we follow an extension of a convection system with an additional model parameter and consider the extended model with a periodic boundary condition
\cite{borue_1997,lohse_2003,calzavarini_2006,liu_fixed-flux_2024,hiruta_simple_2022}.

Bistability is associated with the global structure of the dynamical system, and in particular, an unstable saddle solution is a key structure that characterizes its topological features \cite{itano_dynamics_2001,nishiura_2003}.
Since a saddle solution possesses an attractive direction, orbits in the neighborhood of its stable manifold are attracted towards the saddle point.
Then these orbits are eventually separated along its unstable manifold, forming a boundary set for basins of attraction, called a basin boundary.
Therefore, a saddle solution embedded in the basin boundary fully characterizes the global topological features of the dynamical system and is thus often called an edge state \cite{itano_dynamics_2001,avila_2023}. The importance of an edge state is therefore widely recognized in such bistable systems as is
studied in the Gray-Scott model of collision-division processes \cite{nishiura_2003}, 
a buckling problem of elastic shells \cite{Kreilos_2017}, and bursting phenomena of fluid turbulence \cite{itano_dynamics_2001,Khapko_2013}.

Our primary aim of this paper is therefore to understand the mechanisms of the emergence of the localized stable bioconvection pattern and the bistability at the onset of the convection transition. In doing so, we analyze the stability of the extended model bioconvection system in a periodic boundary condition by introducing the equilibrium density profile as an independent parameter. The secondary aim is then to explore the edge state that characterizes the global structure of the bi-stable dynamical systems.

The rest of this paper is as follows. In Sec.\ref{sec:setting}, we introduce our model system and discuss the key nonlinear coupling between the fluid and density fields. Since the system is dominated by large-scale modes in the vertical direction, we can introduce a simplified system by truncating the higher vertical modes.  In Sec.\ref{sec:linear}, we examine the linear stability of the truncated system and derive the neutrally stable curve. We then provide nonlinear stability analysis, showing that the particle self-propulsion gives rise to a localized convection pattern and system bistability in Sec.\ref{sec:nonlinear}. The emergence of an edge state is also discussed before discussion and conclusions in Sec.\ref{sec:conclusion}.

\section{Problem settings\label{sec:setting}}

\subsection{Model}

We consider a two-dimensional model convection system where the velocity field
$\bm{u}(x,y,t)=(u_x,u_y)$ is coupled with the density deviation $m(x,y,t)$ from an equilibrium density profile $m_0(y)=\sin(y)$
in a doubly periodic domain $(x,y)\in \Omega \equiv [0,L_x] \times [0,2\pi]$. 
We assume that $\bm{u}$ is incompressible ($\nabla \cdot \bm{u}=0$)
and thus the velocity field $\bm{u}$ is  represented by a stream function  $\Psi(x,y,t)$: $(u_x,u_y)=(\partial_y\Psi,-\partial_x \Psi)$.

As an extended model of fluid convection, we consider the following nondimensional equations 
for $m$ and the vorticity $\omega(x,y,t)=\partial_xu_y-\partial_y u_x$:
  \begin{align}
  \frac{\partial\omega}{\partial t}-J(\Psi,\omega)&=Pr( \Delta \omega -Ra \partial_x m) , \label{eq:vorticity2}\\
  \frac{\partial m}{\partial t}-J(\Psi,m)&=\Delta m -Pe \partial_y m +\partial_x \Psi \partial_y m_{0}, \label{eq:density2}
\end{align}
where $J(\Psi,f)$ ($f$ is $\omega$ or $m$) is defined as
\begin{equation}
  J(\Psi,f)\equiv (\partial_x \Psi) (\partial_y f) -(\partial_x f)(\partial_y \Psi).   \label{eq:J}
\end{equation}
Derivations of the model with its physical background are provided in the next subsection [Sec.\ref{app:model}].
Here, three nondimensional parameters, $Ra, Pe, Pr$ represent the Rayleigh number, Prandtl number $Pr$, and P\'eclet number, respectively. 
The P\'eclet number represents nondimensional particle self-propulsive velocity, and when the P\'eclet number approaches zero, Eqs.\eqref{eq:vorticity2}-\eqref{eq:density2} recover the Rayleigh-B\'enard thermal convection, in which the last term in Eq.\eqref{eq:density2} is simply reduced to $\partial_x \Psi$, because of the linear profile of the equilibrium density.
The stream function $\Psi$ satisfies the Poisson equation on the doubly periodic domain $(x,y)\in \Omega$ as follows:
  \begin{align}
    -\Delta\Psi&=\omega. \label{eq:poisson}
\end{align}
We readily find that Eqs.\eqref{eq:vorticity2}-\eqref{eq:density2} have a quiescent state as a trivial solution, $\omega = m = 0$, which 
is a steady solution for any $(Ra,Pr,Pe)$.

\subsection{Derivation of the physical model}
\label{app:model}
In this subsection, we derive our physical model, Eqs. \eqref{eq:vorticity2}-\eqref{eq:density2}, and discuss the underlying physical backgrounds. We follow a standard bioconvection model for a bulk motion\cite{pedley_1992,taheri_bioconvection_2007,bees_2020}, but with a doubly periodic boundary condition. 
By considering an equilibrium density profile as an independent model parameter, we naturally extend the usual bioconvection model to a general boundary condition.

Following well-known continuum description for a dilute solution of self-propelled particles,
we introduce the density field $\tilde{m}(x,y,t)$ and the velocity field for the dilute solution $\bm{u}(x,y,t)=(u_x(x,y,t),u_y(x,y,t))$, where we assume the incomprehensibility condition, $\partial_xu_x+\partial_yu_y=0$.
As the simplest case of bioconvection, we follow Fick's law with a constant diffusion for the density flux, $\bm{J}$, as 
\begin{align}
    J(\tilde{m})&= \tilde{m}\bm{u}+\tilde{m}\bm{V}-D\bm{\nabla} \tilde{m}.
\end{align}
Here, we introduce upward locomotion of the self-propelled motion with $\bm{V}=V\bm{\hat{y}}$, where $\bm{\hat{y}}$ is the unit vector in the $y$ direction and $V$ and $D$ are constant values. The upward motion of self-propelled particles is considered to be essential for bioconvective pattern, and in experiments these biased motions usually originate from the cellular tactic behaviors, e.g. phototaxis, gyrotaxis, and chemotaxis.

Then, we define an equilibrium density profile $m_0(y)$ as the steady solution with $\bm{u}=0$. 
We then introduce the density deviation from the equilibrium profile as $m=\tilde{m}-m_0$.
The density flux for $m$ is then obtained by 
$\delta J\equiv J(\tilde{m})-J(m_0)$, and we have
\begin{align}
    \delta J&= m\bm{u}+mV\bm{\hat{y}}+m_0\partial_yu_y  -D\bm{\nabla} m.
\end{align}
Hence, the time evolution of $m$ follows the continuum equation, $\partial_t m =-\bm{ \nabla} \cdot \delta J$, which is explicitly given by
\begin{align}
    \frac{\partial m}{\partial t}+\bm{u}\cdot\bm{\nabla} m&= -V\partial_y m-u_y\partial_ym_0  +D\Delta m,
\end{align}
where $\Delta = \partial_x^2+\partial_y^2$ is Laplacian in two dimensions.

For the fluid motion, we consider the two-dimensional Navier-Stokes equation with Boussinesq approximation. The velocity $\bm{u}(x,y,t)$ and the pressure $p(x,y,t)$ then satisfy 
\begin{align}
    \rho_s\left(\frac{\partial \bm{u}}{\partial t}+\bm{u}\cdot\bm{\nabla} \bm{u}\right)&= -\bm{\nabla} p+\mu\Delta \bm{u} -  m v (\rho_p-\rho_s) g\bm{\hat{y}}.
\end{align}
Here, $\mu$ is the dynamic viscosity and assumed to be a constant. $\rho_s$ and $\rho_p$ 
are the density of solvent and particles, respectively, and in the context of bioconvection, we usually consider $\rho_p>\rho_s$.
The gravity constant $g>0$ and the particle volume $v$ are also assumed to be constant.

We finally derive our equation of motions Eqs.\eqref{eq:vorticity2}-\eqref{eq:density2}
by introducing the stream function $\Psi$.
Three nondimensional parameters, Rayleigh number $Ra$, Peclet number $Pe$ and Prandtl number $Pr$,
are then introduced with physical parameters as follows:
\begin{align}
Ra &=  \frac{ ||m_0||_{\infty} v(\rho_p-\rho_s)g L_y^3}{8\pi^3 D \mu},\\
Pe &=  \frac{L_y V}{2\pi D},\\
Pr &=  \frac{\mu}{\rho D},
\end{align}
where $L_y$ is the length of domain in $y$.
We have assumed the length scale $L=L_y/2\pi$, the time scale $T=L^2/D$, and the density scale
as $||m_0||_{\infty}$.
Our model includes 
Rayleigh--B\'enard convection, which is the most fundamental model for thermal convection,
as the case for $Pe=0$ and $dm_0/dy=$constant. If a non-slip or free-slip boundary condition is imposed on the bottom and top boundaries of the fluid region, the equilibrium density profile is then accordingly determined. In this sense, our extended model captures the bulk behaviour of bioconvection.

\subsection{Truncation of vertical modes}
Focusing on horizontal accumulation of particles, we introduce vertically averaged density $n(x,t)\equiv\langle m \rangle$, where the bracket $\langle \rangle$ is
average in $y$, defined as
  \begin{align}
    \langle \alpha \rangle\equiv \frac{1}{2\pi}\int_0^{2\pi} \alpha(x,y,t) dy. \label{eq:n0}
\end{align}
for a field variable $\alpha(x,y,t)$.
By integrating Eq.\eqref{eq:density2} in the $y$ direction, we find
  \begin{align}
    \frac{\partial n}{\partial t}&=\partial_x^2 n +\partial_x \langle (m_{0}+m)u_x \rangle. \label{eq:density0}
\end{align}
This equation suggests that the particle accumulation can occur only when the second term of Eq.\eqref{eq:density0} has a nonzero contribution.

\begin{figure}
   \begin{overpic}[width=0.28\linewidth]{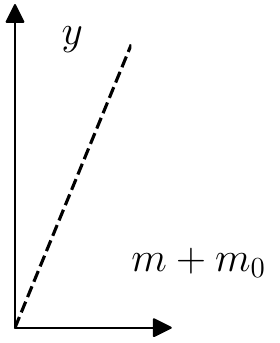}
    \put(-6,102){(a)}
    \end{overpic}
   \begin{overpic}[width=0.48\linewidth]{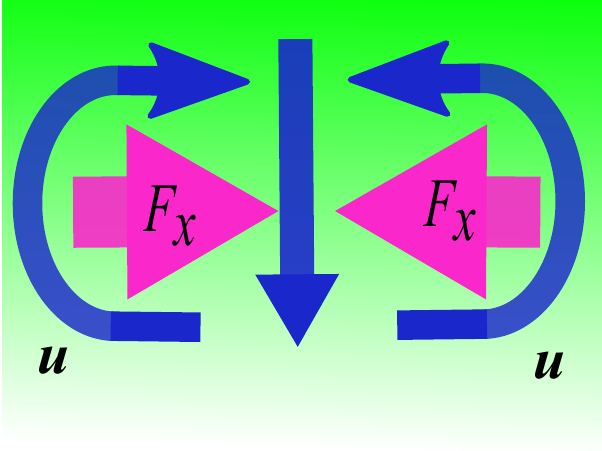}
    \put(-10,73){(b)}
    \end{overpic}    
  \caption{\label{fig:idea} Schematic of nonlinear couplings between the fluid flow and the fluid density that could generate a large-scale circular convection pattern. (a) We assume that the density profile monotonically increases in the vertical direction. (b) The base density profile is shown in the green layer. When spatial accumulation of particle density occurs, the net flux of density, $F_x\equiv-\langle (m_0+m)u_x\rangle$, is generated as depicted in the magenta arrows. This incoming density flux is accompanied by the downward jet and a large-scale circulatory fluid motion. See the main text for details.
  }
\end{figure}

To physically interpret this nonlinear effect, we introduce a horizontal net density flux as $F_x\equiv-\langle (m_0+m)u_x\rangle$. As shown in Fig.\ref{fig:idea}(a), let us consider the situation where the particle concentration is denser in the upper region [also illustrated by the green shades in Fig. \ref{fig:idea}(b)]. To generate the localized convection pattern from the horizontally homogeneous trivial state, nonzero horizontal net flux must be produced as indicated by magenta arrows in Fig.\ref{fig:idea}(b). Since the sign of $F_x$ is dominated by the sign of $u_x$ in the region of higher $m+m_0$ values, the horizontal fluid velocity $u_x$ needs to be generated in the same direction as the net density flux at the upper region, leading to an incoming flow at the point of the particle condensation and the associated  outgoing flow at the lower region due to the fluid incompressibility. In the end, these mechanical couplings result in the downward jet accompanied by the particle condensation with a circulatory fluid motion as in the blue arrows in Fig.\ref{fig:idea}(b).

These physical mechanisms imply that the trivial solution could be destabilized by the nonlinear coupling and then the system forms a large-scale convective pattern. We therefore focus on the large-scale coupling to capture the dominant nonlinear effects of the system. In doing so, we introduce a minimal system by truncating the higher Fourier modes in the vertical direction. We retain only the lowest three Fourier modes which are proportional to $\exp(ily)$ with $l=0,\pm1$ as
\begin{align}
  \omega(x,y,t) & = A_{-1}(x,t) \exp(-iy)+A_{0}(x,t) +A_{1}(x,t) \exp(iy),  \label{eq:trunc1} \\
  m(x,y,t)      & = B_{-1}(x,t) \exp(-iy)+B_{0}(x,t) +B_{1}(x,t) \exp(iy),   \label{eq:trunc2}\\
  \Psi(x,y,t)      & = C_{-1}(x,t) \exp(-iy)+C_{0}(x,t) +C_{1}(x,t) \exp(iy),  \label{eq:trunc3}
\end{align}
where $i=\sqrt{-1}$ is the imaginary unit.
To validate our truncation model, we have also numerically confirmed that the qualitative behaviors of the system are unchanged even in the presence of the higher modes, $l=0,\pm 1,\cdots, \pm 64$. Henceforth, in this paper, we investigate the minimal large-scale model with Eqs. \eqref{eq:trunc1}-\eqref{eq:trunc3}.

Since $\omega$ and $m$ are variables with real value, $A_{0}$ and $B_0$ must also be real.
Similarly, $A_{1}^{\ast}=A_{-1}$ and $B_{1}^{\ast}=B_{-1}$ hold, where the asterisk indicates complex conjugate. From the Poisson equation \eqref{eq:poisson}, the stream function $\Psi$ satisfies
  \begin{align}
    -(\partial_x^2-l^2)C_{l}&=A_{l}, \label{eq:poisson2}
\end{align}
for $l=0$, and $1$.

\subsection{Numerical methods \label{sec:numerical}}

We first expand $A_{l}$, $B_{l}$, and $C_{l}$ ($l=0, \pm1$) by horizontal Fourier series as 
  \begin{align}
    A_{l}(x,t)&=\sum_{k=-M}^{M} \tilde{A}_{k,l}(t)\exp(ikx) \label{eq:expA},\\
    B_{l}(x,t)&=\sum_{k=-M}^{M} \tilde{B}_{k,l}(t)\exp(ikx) \label{eq:expB},\\
    C_{l}(x,t)&=\sum_{k=-M}^{M} \tilde{C}_{k,l}(t)\exp(ikx) \label{eq:expC},
\end{align}
where we choose a truncation wavenumber as $M=64$ to sufficiently resolve horizontal structures.
Hence, our model equation, Eqs. \eqref{eq:trunc1} - \eqref{eq:trunc3}, are discretized into the following equations:
  \begin{align}
    \left(\frac{d}{dt}+Pr k^2\right) \tilde{A}_{k,0} &=-ikPrRa  \tilde{B}_{k,0} 
    +\sum_{p,q}qk\frac{\tilde{A}_{p,q}\tilde{A}_{k-p,-q}}{p^2+q^2}, 
   \label{eq:g1}\\
    \left(\frac{d}{dt}+Pr (k^2+1)\right)\tilde{A}_{k,1}&= -ikPrRa \tilde{B}_{k,1}
    +\sum_{p,q}(qk-p)\frac{\tilde{A}_{p,q}\tilde{A}_{k-p,1-q}}{p^2+q^2}, \\
  \left(\frac{d}{dt}+k^2\right)\tilde{B}_{k,0}
  &=ik\frac{\tilde{A}_{k,1}+\tilde{A}_{k,-1}}{2(k^2+1)}
  +\sum_{p,q}qk\frac{\tilde{A}_{p,q}\tilde{B}_{k-p,-q}}{p^2+q^2}, \label{eq:g3}\\
  \left(\frac{d}{dt}+(k^2+1)+iPe\right)\tilde{B}_{k,1}
  &=ik\frac{\tilde{A}_{k,0}}{k^2}+\sum_{p,q}(qk-p)\frac{\tilde{A}_{p,q}\tilde{B}_{k-p,1-q}}{p^2+q^2}. \label{eq:g4}
\end{align}
Here, we used the relation, $C_{k,l}=(k^2+l^2)^{-1}A_{k,l}$, which follows from Eq.\eqref{eq:poisson2}.

Notably, the system has a reflection symmetry in $x$, as Eqs.\eqref{eq:vorticity2} and \eqref{eq:density2} are unchanged
by a reflection $R_{x}:(\omega(x,y,t),m(x,y,t))\rightarrow(-\omega(-x,y,t),m(x,y,t))$, which yields the relation between the horizontal Fourier coefficients
$\tilde{A}_{k,l}=-\tilde{A}_{-k,l}$ and $\tilde{B}_{k,l}=\tilde{B}_{-k,l}$.
Since the real condition of $\omega$ and $m$ read 
$\tilde{A}_{k,l}=\tilde{A}_{-k,-l}^{\ast}$ and 
$\tilde{B}_{k,l}=\tilde{B}_{-k,-l}^{\ast}$,
we obtain the relation,
$\tilde{A}_{k,-l}=-\tilde{A}^{\ast}_{k,l}$ and $\tilde{B}_{k,-l}=\tilde{B}_{k,l}^{\ast}$, which, for example, guarantees that the first term in the right-hand side of Eq.\eqref{eq:g3} is real.   

By this symmetry, $\tilde{A}_{k,0}$ and $\tilde{B}_{k,0}$ are pure imaginary and real numbers, respectively, and the time evolution of our system, Eqs. \eqref{eq:g1}-\eqref{eq:g4}, is determined by a set of Fourier modes with non-negative $k$ indices, leading to a dynamical system of a state $\bm{X}(t)$ with $(M+1)\times 6$ real variables.

In our numerical computations, we have utilized the pseudospectral method for computing nonlinear terms
complemented by a Fast Fourier Transform (FFT).
In time stepping, the linear terms have been dealt as the integral factor and we have solved the nonlinear terms by the 4th-order Runge-Kutta method.

To compute stationary solutions,
we have applied the GMRES-Newton method, in which the linear space is approximated in a Krylov subspace.
Iterations of Newton's method have been terminated when the time derivatives of the Fourier modes are sufficiently small. More precisely, we employed the condition, 
$||d\bm{X}/dt||_2<10^{-5}$, where $|| \bullet||_2$ indicates the Euclidean norm.
The stationary solutions in different parameters are continuously obtained by the arclength continuation method.

\section{Linear stability of trivial solution\label{sec:linear}}

In this section, we theoretically examine the linear stability of the trivial solution and its dependence on $Pe$.

We consider a sinusoidal disturbance to the trivial steady state, $\omega=m=0$, with a wavenumber $k=k_0$, assuming the solution of the forms $\omega=e^{\lambda t+ik_0 x}\left(\tilde{A}_{k_0,-1}e^{-iy}+\tilde{A}_{k_0,0}+\tilde{A}_{k_0,1}e^{iy})\right)$ and $m=e^{\lambda t+ik_0 x}\left(\tilde{B}_{k_0,-1}e^{-iy}+\tilde{B}_{k_0,0}+\tilde{B}_{k_0,1}e^{iy})\right)$.
  Accordingly, the linear stability analysis is reduced to an eigenvalue problem, 
  $\lambda \bm{x}_{\lambda} = M \bm{x}_{\lambda}$,
  where $\bm{x}$ is the 6-dimensional vector, 
  $\bm{x}=(\tilde{A}_{k_0,1},\tilde{A}_{k_0,0},\tilde{A}_{k_0,-1},  \tilde{B}_{k_0,1},\tilde{B}_{k_0,0},\tilde{B}_{k_0,-1})^{T}$, and the $6\times 6$ matrix $M$ is given by\\
  \begin{equation}
  M=
  \begin{pmatrix}
  -Pr(k_0^2+1) & 0 & 0&  -ik_0PrRa & 0 & 0 \\
  0 & -Prk_0^2 & 0    &  0 & -ik_0PrRa & 0 \\
  0 & 0 & -Pr(k_0^2+1) &   0 & 0 & -ik_0PrRa \\
  0 & \frac{i}{2k_0} & 0 &                      -(k_0^2+1) -iPe& 0 & 0 \\ 
  \frac{ik_0}{2(k_0^2+1)} & 0 & \frac{ik_0}{2(k_0^2+1)}&  0 & -k_0^2 & 0 \\ 
  0 & \frac{i}{2k_0} & 0                    &      0 & 0 & -(k_0^2+1) +iPe
  \end{pmatrix}.
  \end{equation}

The eigenvalue $\lambda$ follows the characteristic equation of the six degrees, 
\begin{equation}
    0  = \frac{\left(Pr (k_0^{2} +1) + \lambda\right)}{k_0^{2} + 1} 
    \left(a_5 \lambda^5 +a_4\lambda^4+a_3\lambda^3 +a_2\lambda^2 +a_1\lambda  +a_0\right),
    \label{eq:char_eq}
\end{equation}
where the coefficients are explicitly given by
\begin{eqnarray*}
    a_5 &\equiv & k_0^{2}+1 \\
    a_4 &\equiv &\left(k_0^{2} +1\right) \bigg(2 Pr k_0^{2} + Pr  + 3 k_0^{2} + 2 l^{2}\bigg) \\
    a_3 &\equiv &\left(k_0^{2} + 1\right) \bigg(Pe^{2} + Pr^{2} k_0^{4} + Pr^{2} k_0^{2}
     + 6 Pr k_0^{4} + 7 Pr k_0^{2} 
    + 2 Pr + 3 k_0^{4} + 4 k_0^{2}  + 1\bigg)\\
    a_2&\equiv&\left(k_0^{2} + 1\right) \bigg(2 Pe^{2} Pr k_0^{2} + Pe^{2} Pr  + Pe^{2} k_0^{2} + 3 Pr^{2} k_0^{6}
    + 5 Pr^{2} k_0^{4} + 2 Pr^{2} k_0^{2}  + 6 Pr k_0^{6} + 11 Pr k_0^{4} 
    + 6 Pr k_0^{2}  + Pr  + k_0^{6} + 2 k_0^{4}  + k_0^{2} \bigg)\\
    a_1 &\equiv& Pr k_0^{2} \bigg( Pe^{2} Pr k_0^{4} + 2 Pe^{2} Pr k_0^{2}  + Pe^{2} Pr  + 2 Pe^{2} k_0^{4}
    + 3 Pe^{2} k_0^{2} + Pe^{2} \\
    &&-  \frac{Ra^{2} Pr}{2} + 3 Pr k_0^{8} + 10 Pr k_0^{6} 
    + 12 Pr k_0^{4} + 6 Pr k_0^{2} + Pr + 2 k_0^{8} + 7 k_0^{6}  + 9 k_0^{4}  + 5 k_0^{2}  + 1\bigg)\\
    a_0&\equiv& Pr^{2} k_0^{2} \left(k_0^{2} + l^{2}\right) 
    \bigg( Pe^{2} k_0^{4} + Pe^{2} k_0^{2}  - \frac{ Ra^{2}}{2} + k_0^{8} + 3 k_0^{6}  + 3 k_0^{4}  + k_0^{2} \bigg).
\end{eqnarray*}

  The sixth-degree equation \eqref{eq:char_eq} contains a trivial root of $\lambda=-Pr (k_0^2+1)$, and the other five roots are those of the quintic equation, 
    $f(\lambda) = a_5 \lambda^5+a_4 \lambda^4+a_3 \lambda^3+a_2 \lambda^2+a_1 \lambda +a_0=0$.
    We first numerically confirmed in a parameter region of our interest that an eigenvalue with a positive real part is a real number, indicating that the Hopf bifurcation does not occur in this system. We therefore focus on the real eigenvalues and examine the neutral stability curve. 
    Although there exists no algebraic method to find roots, we can formally argue the linear stability via the following proposition:
  \begin{prop}
    There exists only one positive real root of $f(\lambda)=0$ if $a_0<0$, and if $a_0>0$, the real roots of $f(\lambda)=0$ are all negative. If $a_0=0$, $\lambda=0$ is a root of $f(\lambda)=0$ and other real roots are negative. 
  \end{prop}
  One can readily prove this proposition by observing the following two properties of the coefficients:
  (1) $a_5$, $a_4$, $a_3$ and $a_2$ are positive, and (2) $a_1$ is positive whenever $a_0$ is positive. The latter readily follows when we rewrite $a_1$ as
  \begin{eqnarray}
  a_1&=&\frac{a_0}{(k_0^2+1)} +Prk_0^2(k_0^2+1)\bigg[Pe^2(Pr+2k_0^2+1)+(Pr+1)(k_0^2+1)^2(2k_0^2+1)\bigg].
  \end{eqnarray}
  The zeros of $f(\lambda)$ are the intersection of the following two functions: $f_1(\lambda)=-a_5\lambda^5-a_4\lambda^4-a_3\lambda^3-a_2\lambda^2$ and $f_2(\lambda)=a_1\lambda +a_0$. Here, $f_1(\lambda)$ is a monotonically decreasing function of $\lambda$ and satisfies $f_1(0)=0$.  If $a_0>0$ and $\lambda>0$, we find $f_1(\lambda)<0$ and $f_2(\lambda)>0$, the latter of which follows from property (2).
  Hence, these two functions do not have an intersection, resulting in no zeros with a positive real part for $f(\lambda)$ if $a_0>0$. 
  If $a_0$ is negative ($a_0<0)$, on the other hand, we have $f_1(0)=0$ and $f_2(0)<0$ while inequality $f_1(\lambda)<f_2(\lambda)$ should hold at a sufficiently large value of $\lambda>0$, indicating that there is at least one positive zero of $f(\lambda)$ according to the intermediate value theorem. Since $f_1(\lambda)$ is upper-convex on $\lambda>0$, there exists only one positive real root for $f(\lambda)=0$. The nonexistence of roots with positive real parts at $a_0=0$ follows with a similar argument.\\

Hence, the sign of $a_0$ determines the linear stability of the trivial solution, and the neutral stability plane is obtained as the hyperplane of $f(0) = a_0 =0$ in the parameter space $(Ra,Pe,k_0)$, or equivalently,
  \begin{align}
    N(Ra,Pe,k_0)&\equiv\frac{Ra^2}{2}-Pe^2k_0^2(k_0^2+1)-k_0^2(k_0^2+1)^3=0. \label{eq:N}
  \end{align}
Since $N$ is a monotonically decreasing function of $k_0$,
the system is linearly stable against any disturbance when $N(Ra,Pe,k_s)<0$, where $k_s\equiv2\pi/L_x$ is the smallest wavenumber determined by the system size.
 We also find that the trivial solution is stabilized as $Pe$ increases and that the stability condition is independent of $Pr$. The critical P\'eclet number, below which the system becomes linearly unstable for given values of $Ra$ and $L_x$, is explicitly given by  
  \begin{align}
    Pe^{c}(Ra,k_s) = \sqrt{\max \left(0,\frac{Ra^2}{2(k_s^2+1)k_s^2} -(k_s^2+1)^2 \right) }.\label{eq:Pec}
  \end{align}

\section{Bifurcation of nonlinear steady solutions\label{sec:nonlinear}}
\subsection{Bifurcation diagram}

We then proceed to analyze the nonlinear regime, focusing on smaller P\'eclet numbers where the trivial solution becomes linearly unstable. 
To evaluate the impact of $Pe$ on the convection structure and its stability, we performed a
bifurcation analysis with the horizontal scale of the fluid region fixed as $L_x=8\pi$, following the existing literature on the thermal convection and forced Navier-Stokes flow problems in a doubly periodic domain \cite{taheri_bioconvection_2007, lucas_spatiotemporal_2014,hiruta_solitary_2015,hiruta_intermittent_2017}. 
The critical Rayleigh number at $Pe=0$ is then obtained from Eq. \eqref{eq:Pec} as $Ra=\sqrt{2k_s^2(1+k_s^2)^3}\approx 0.388$. Since our current focus is on the nonlinear regime, we choose the Rayleigh number to be larger than this critical value and set $Ra = 0.5$. We also set the Prandtl number $Pr=2.5$ as a reasonable value in the experimental setup of bioconvection of {\it Euglena} \cite{giometto_2015}. This choice of value is also reasonable in study of {\it Paramecium} by Taheri \& Bilgen \cite{taheri_bioconvection_2007} in which the nondimensional parameter ranges are estimated as $Pr=0.22-2$ and $Pe=0.07-1.54$.

\begin{figure}
    \begin{minipage}{0.45\linewidth}
   \begin{overpic}[width=\linewidth]{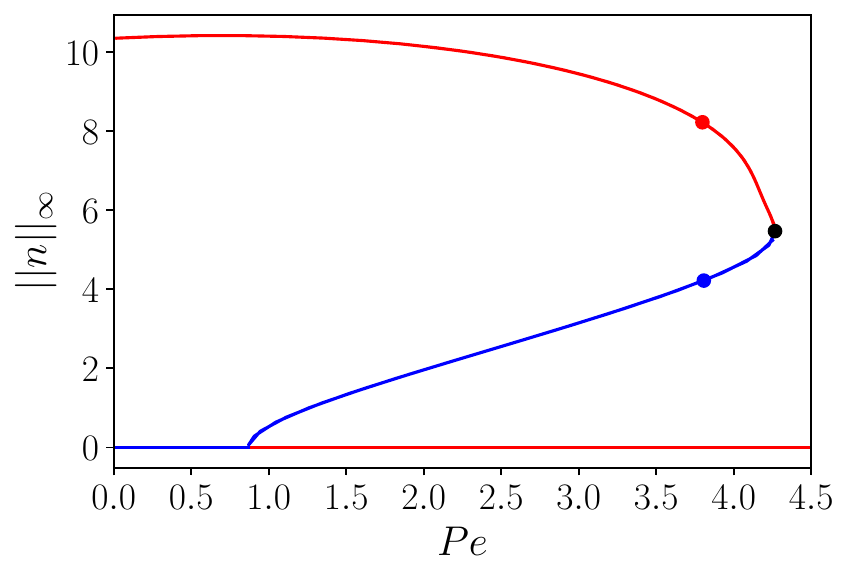}
    \put(0,65){(a)}
    \end{overpic}
    \end{minipage}
    \begin{minipage}{0.45\linewidth}
   \begin{overpic}[width=\linewidth]{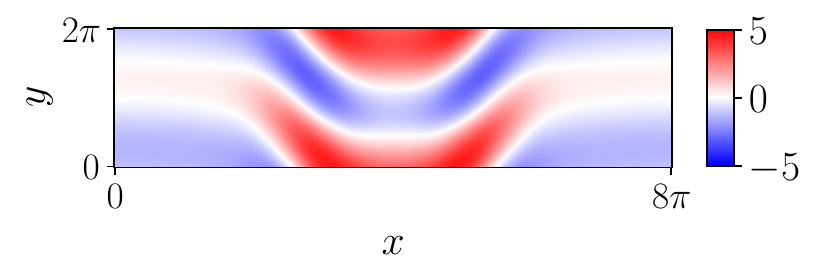}
    \put(0,30){(b)}
    \end{overpic}
   \begin{overpic}[width=\linewidth]{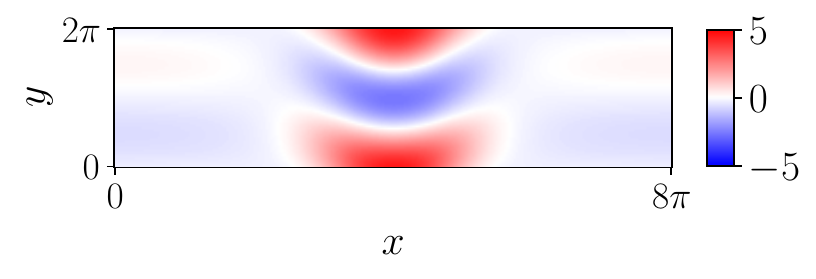}
    \put(0,30){(c)}
    \end{overpic}    
    \end{minipage}
  \caption{\label{fig:bifurcation_sc25}(a) Bifurcation diagram of the steady solution for different values of $Pe$ with $Pr=2.5$ and $Ra=0.5$. The trivial solution becomes unstable below the critical P\'eclet number ($Pe^c\approx0.868)$. Stable and unstable solutions, shown in red and blue, respectively, collapsed at $Pe_{SN}\approx 4.27$, where a saddle-node bifurcation occurs.
  (b) The density deviation field $m(x,y)$ of the stable upper branch at $Pe=3.8$, $\bm{X}_{\rm UB}$ [marked by a red dot in (a)]
  (c) The density deviation field $m(x,y,t)$ of the unstable lower branch at $Pe=3.8$, $\bm{X}_{\rm LB}$ [marked by a blue dot in (a)] 
  } 
\end{figure}

We first show a bifurcation curve in Fig.\ref{fig:bifurcation_sc25}(a), where  the max norm of $n$, $||n||_\infty$, is plotted against the P\'eclet number $Pe$.
The stable and unstable steady solutions are colored red and blue, respectively.
At $Pe=0$, a nonlinear stable solution exists in addition to the unstable trivial solution,
and a unstable nonlinear solution bifurcates from the trivial solution at the critical P\'eclet number, $Pe=Pe_{c}$. The stable and unstable nonlinear solutions collapse at $Pe_{SN}\approx 4.27$, where a saddle-node bifurcation occurs. In the region of $Pe \in (Pe^{c},Pe_{SN}) \approx (0.868,4.27)$, there exist two nonlinear solutions, and we hereafter call the stable solution as UB (upper branch) solution denoted by the state vector $\bm{X}_{\rm UB}$ 
and the unstable solution as LB (lower branch) solution, which we denoted by $\bm{X}_{\rm LB}$.

In Fig.\ref{fig:bifurcation_sc25}(b,c), we show density deviation fields, $m(x,y)$, for the upper and lower branches at $Pe=3.8$, which correspond to the red and blue circles in Fig.\ref{fig:bifurcation_sc25}(a), respectively. In both cases, the density deviation field is spatially localized. Note that the system possesses a horizontal translational symmetry, and we choose a solution that is localized at the middle of the domain. The differences between the two steady solutions are discussed detailed in Sec. \ref{sec:spatial}.

Beyond the saddle-node bifurcation $(Pe>Pe_{SN})$, we numerically confirmed that there is only a stable trivial solution.
Since the P\'eclet number represents the nondimensional self-propulsive velocity, these results suggest that the self-motility could induce the bistable structure in the convection system.

\subsection{Spatial structure of nonlinear steady solutions}
\label{sec:spatial}
\begin{figure}[tb!]
   \begin{overpic}[width=0.45\linewidth]{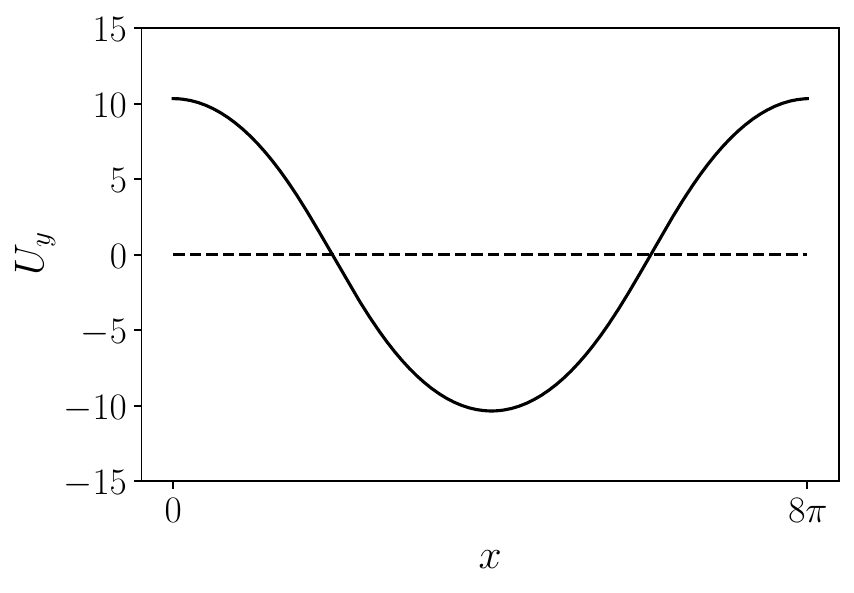}
    \put(0,65){(a)}
    \end{overpic}
    \begin{overpic}[width=0.45\linewidth]{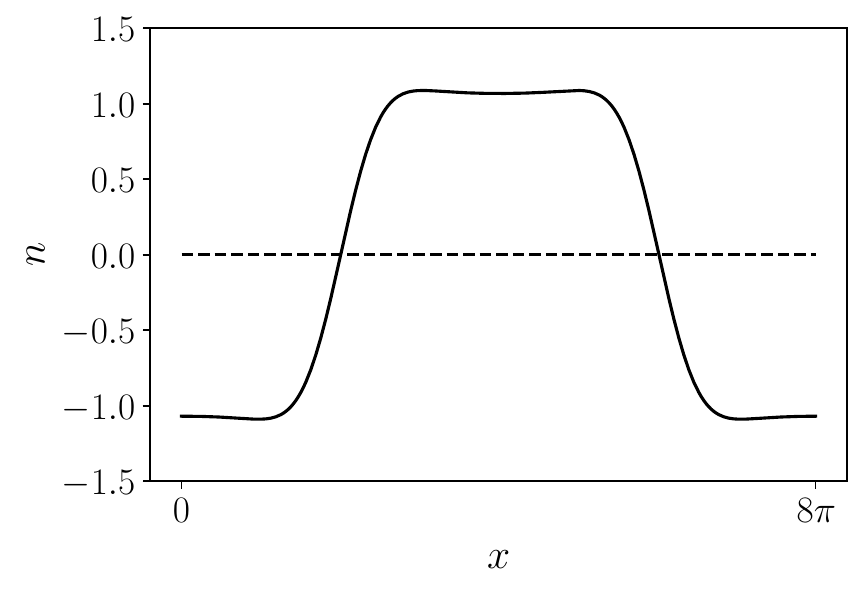}
    \put(0,65){(b)}
    \end{overpic}\\
      \begin{overpic}[width=0.45\linewidth]{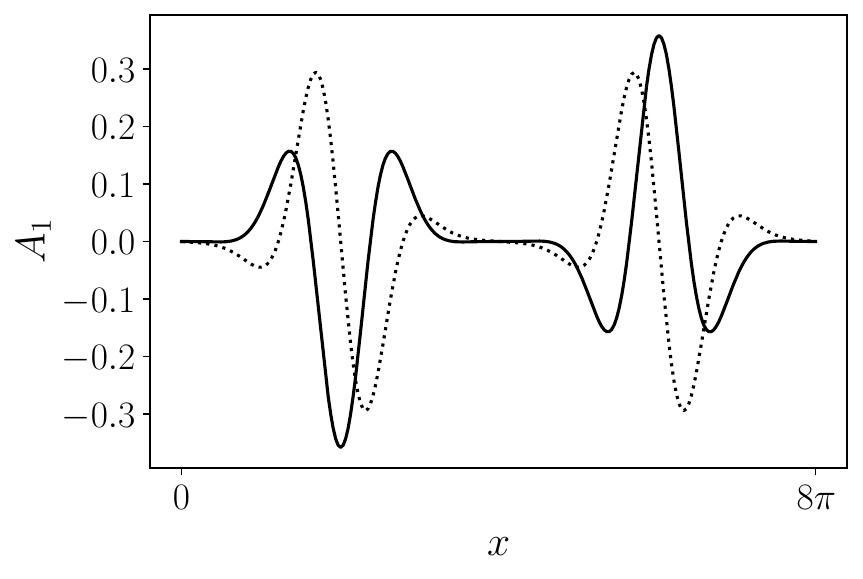}
    \put(0,65){(c)}
    \end{overpic}
      \begin{overpic}[width=0.45\linewidth]{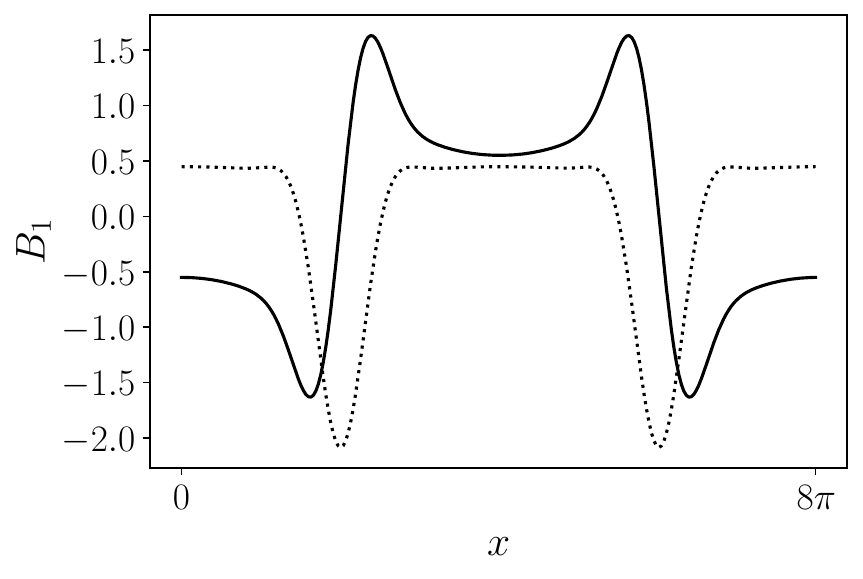}
    \put(0,65){(d)}
    \end{overpic}
  \caption{\label{fig:Pe0} Spatial structure of non-trivial stable solution at $Pe=0$.  
  (a) vertically averaged upward velocity, $U_y(x)$, and (b) the vertically-averaged density deviation $n(x)$. (c) Complex amplitude for the vorticity field, $A_1(x)$, and (d) for the density deviation field, $B_1$.
  The solid lines indicate the real parts and the broken lines indicate the imaginary parts.}
\end{figure}

We then focus on the spatial structure of the fluid and particle density fields. To characterize steady solutions, we consider a vertical average of the upward velocity and a similar vertical average of the density deviation, which are denoted by $U_y\equiv \langle u_y(x,y)\rangle=-\partial_xC_0$ and $n(x)=\langle m(x,y)\rangle$, respectively. We plot these values for the nonlinear stable solution at $Pe=0$ in Fig.\ref{fig:Pe0}(a) and Fig.\ref{fig:Pe0}(b), as well as the two complex modes of the vorticity and density deviation fields [see Eqs. \eqref{eq:trunc1}-\eqref{eq:trunc2}], $A_1(x)$ and $B_1(x)$, in Fig.\ref{fig:Pe0}(c) and Fig.\ref{fig:Pe0}(d). The real and imaginary parts are shown as solid and broken lines, respectively.

From these plots, we find that 
the field structure is horizontally separated into two region; one is characterized by 
the upward fluid velocity and the dilute density field, and the other has the opposite properties. This anticorrelation between the fluid and density fields agrees with the schematics of the nonlinear coupling in Fig.\ref{fig:idea}. Also, this result is consistent with experimental observations of bioconvection \cite{suematsu_2011,shoji_2014,yamashita_2023,bees_2020}.
We observe that the spatial separation is visible more clearly for the density field, as seen in the kink-like structures of the figure. The complex amplitudes in Fig.\ref{fig:Pe0}(c,d) exhibit peaky structures around the kinks seen in the plots of $U_y$ and $n$. 

\begin{figure}[tb!]
   \begin{overpic}[width=0.45\linewidth]{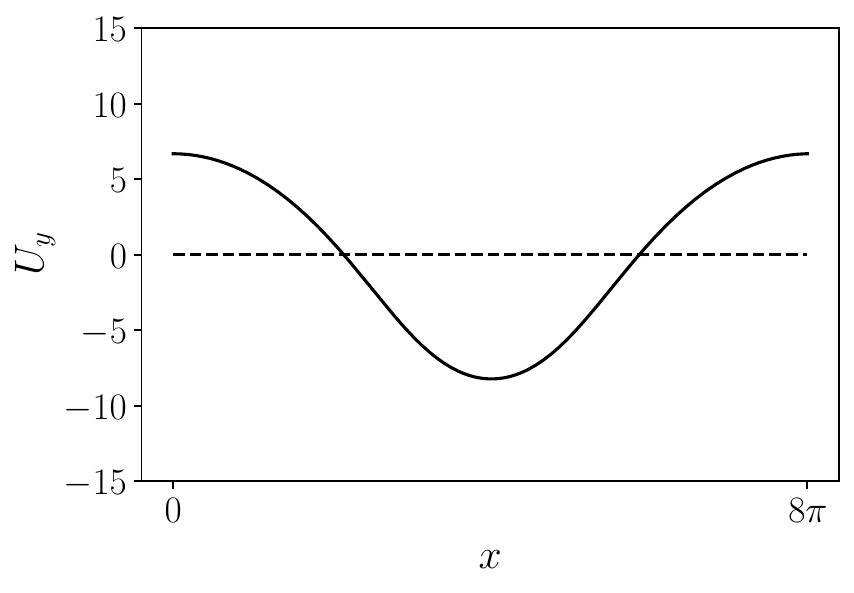}
    \put(0,65){(a)}
    \end{overpic}
   \begin{overpic}[width=0.45\linewidth]{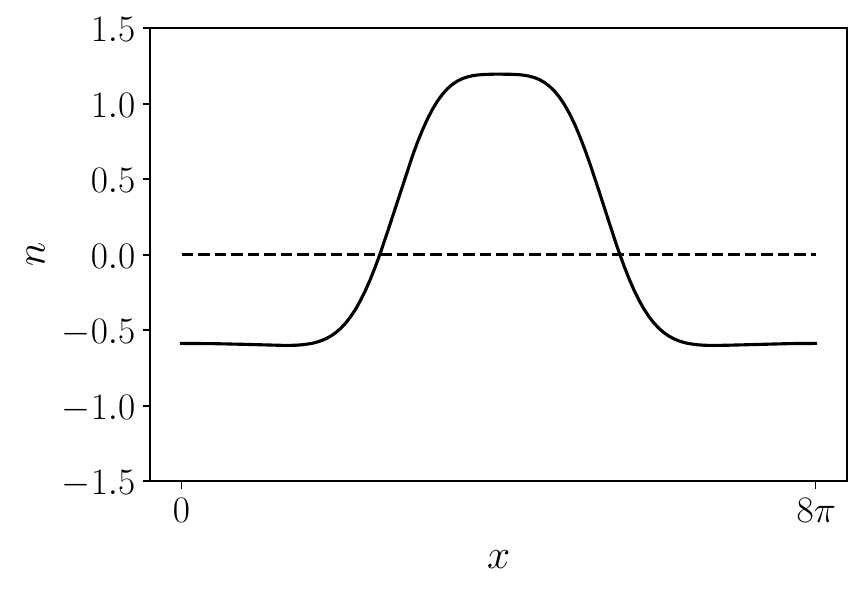}
    \put(0,65){(b)}
    \end{overpic}\\
   \begin{overpic}[width=0.45\linewidth]{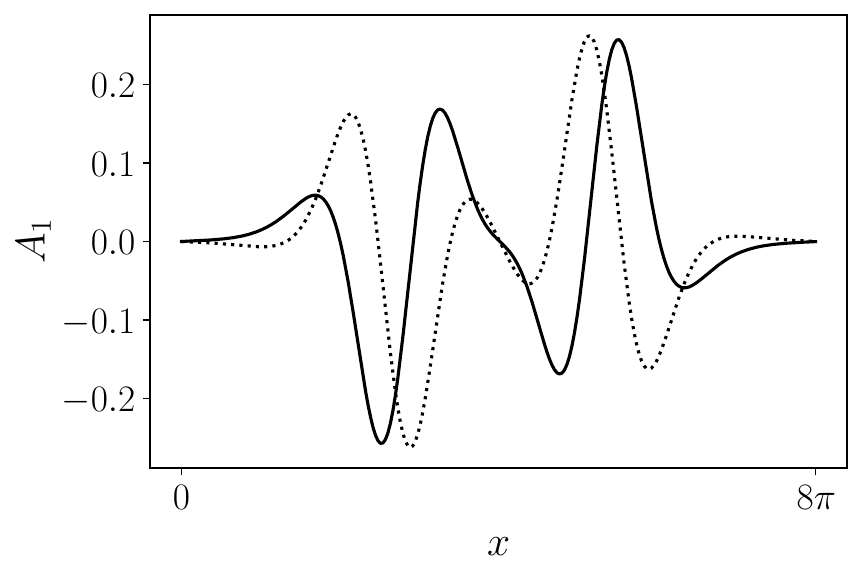}
    \put(0,65){(c)}
    \end{overpic}
   \begin{overpic}[width=0.45\linewidth]{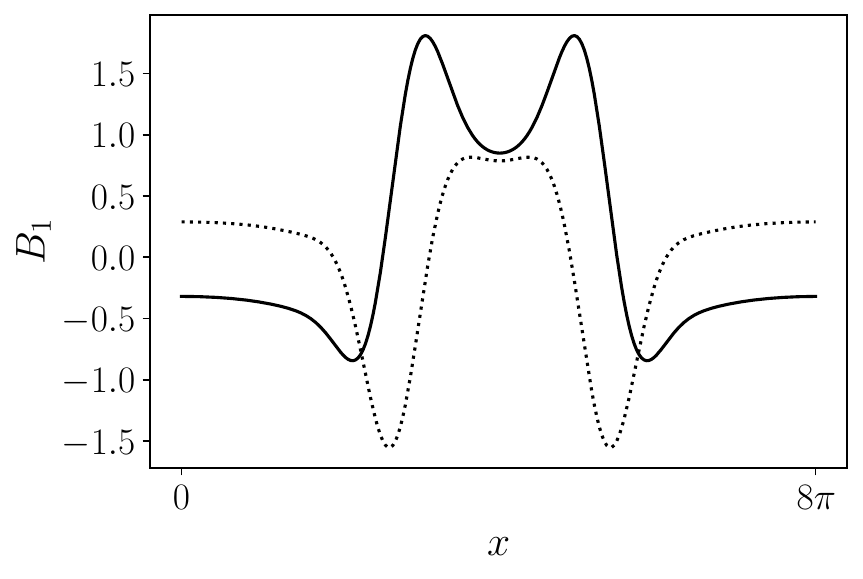}
    \put(0,65){(d)}
    \end{overpic}    
  \caption{\label{fig:UB} Spatial structure of the UB solution, $\bm{X}_{\rm UB}$, at $Pe=3.8$.  
  (a) vertically averaged upward velocity, $U_y(x)$, and (b) the vertically-averaged density deviation $n(x)$. (c) Complex amplitude for the vorticity field, $A_1(x)$, and (d) for the density deviation field, $B_1$.}
\end{figure}

We then examine the same stable nonlinear solution but with an increased P\'eclet number. In Fig.\ref{fig:UB}, we show the similar plots to characterize the spatial structure at $Pe=3.8$.

As in Fig. \ref{fig:Pe0}, we find the large-scale spatial separation with the anticorrelation between the velocity and density fields. Compared with the density concentration at $Pe=0$, the density field $n(x)$ in Fig.\ref{fig:UB}(b) exhibits a further localized structure.
To quantify the width between the kink and anti-kink, we introduce a size of the concentrated region as $L_n=\frac{\min n}{\min n -\max n}L_x$.
By numerically evaluating the values of $L_n$ in different $Pe$, we found that as $Pe$ increases $L_n$ monotonically decreases from $L_n=4\pi=0.5L_x$ at $Pe = 0$ to $L_n\approx 0.204 L_x$ at $Pe=Pe_{\rm SN}$.

\begin{figure}[tb!]
  \begin{overpic}[width=0.45\linewidth]{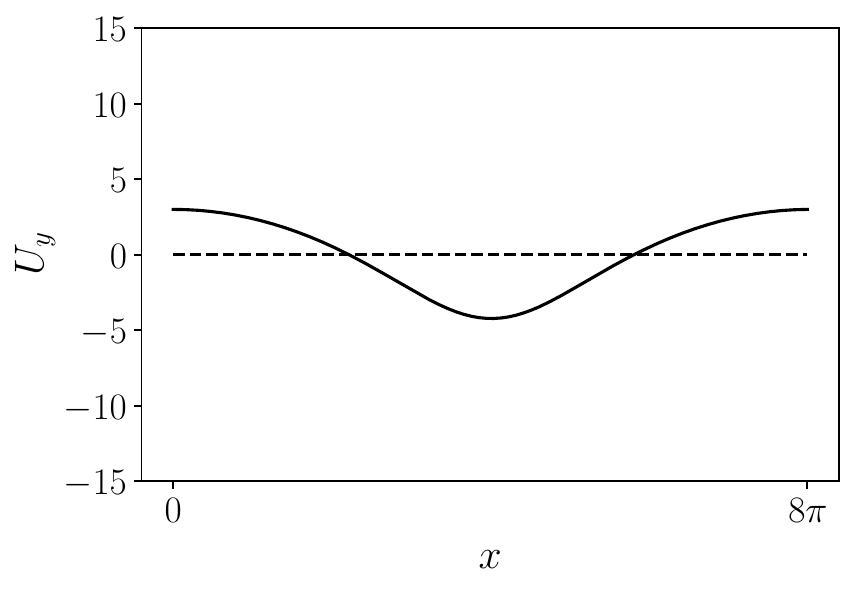}
    \put(0,65){(a)}
    \end{overpic}
  \begin{overpic}[width=0.45\linewidth]{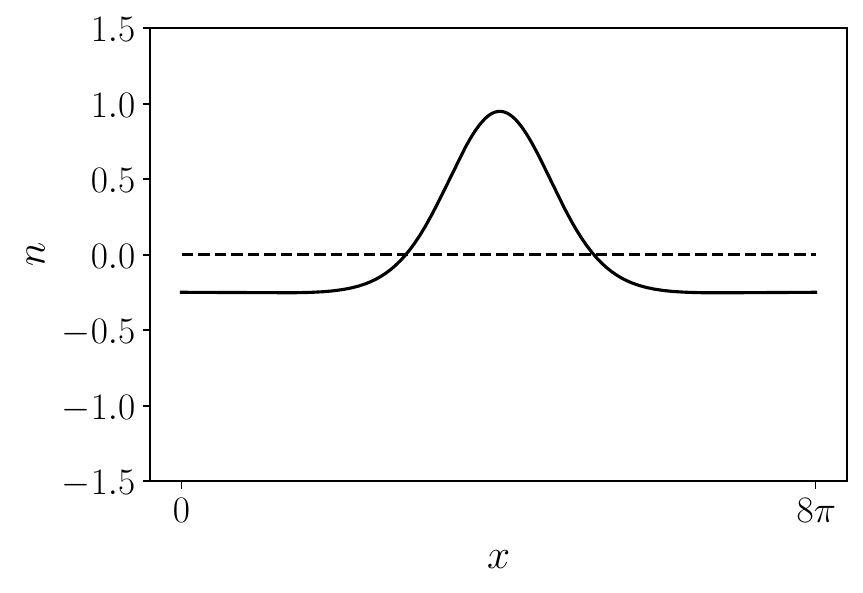}
    \put(0,65){(b)}
    \end{overpic}\\
  \begin{overpic}[width=0.45\linewidth]{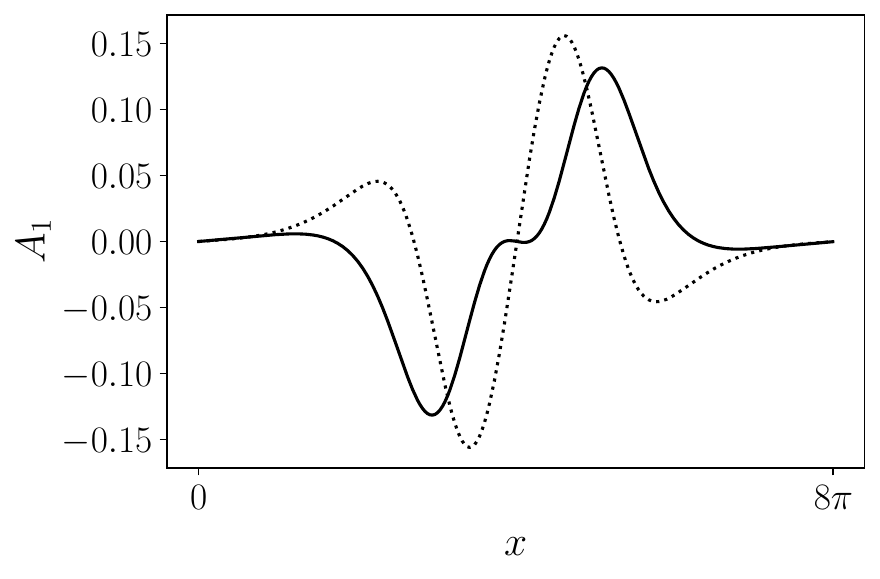}
    \put(0,65){(c)}
    \end{overpic}
  \begin{overpic}[width=0.45\linewidth]{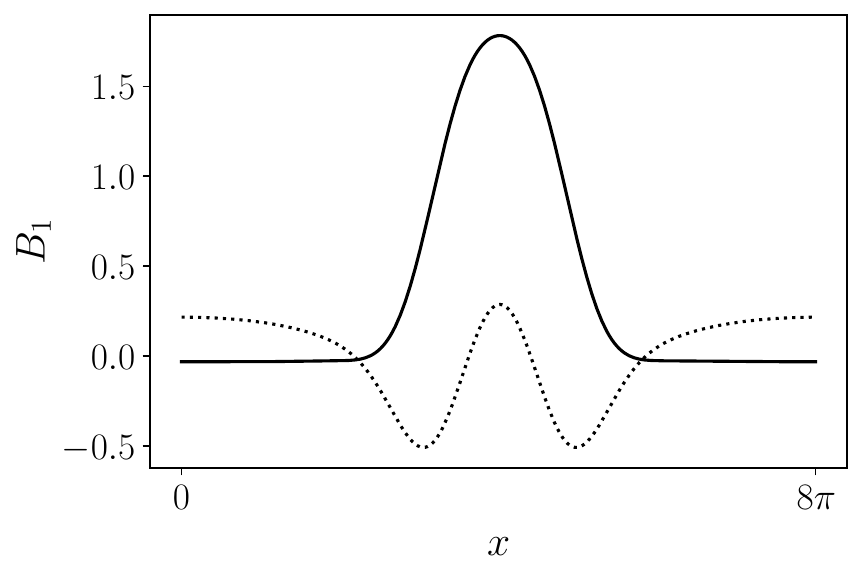}
    \put(0,65){(d)}
    \end{overpic}
  \caption{\label{fig:LB} Spatial structure of the LB solution, $\bm{X}_{\rm LB}$, at $Pe=3.8$.  
  (a) vertically averaged upward velocity, $U_y(x)$, and (b) the vertically-averaged density deviation $n(x)$. (c) Complex amplitude for the vorticity field, $A_1(x)$, and (d) for the density deviation field, $B_1$.}
\end{figure}

We then proceed to study the LB solution, which forms the unstable lower branch in Fig.\ref{fig:bifurcation_sc25}. We examine the spatial structure of $X_{\rm LB}$ at $Pe=3.8$ and the results are summarized in Fig.\ref{fig:LB}. As shown in Fig.\ref{fig:LB}(a,b), the LB solution exhibits spatially-localized profiles both in the vertical velocity and density concentration than the UB solution, although the amplitudes of the profile are slightly suppressed.

\begin{figure}[tb!]
  \begin{overpic}[width=0.45\linewidth]{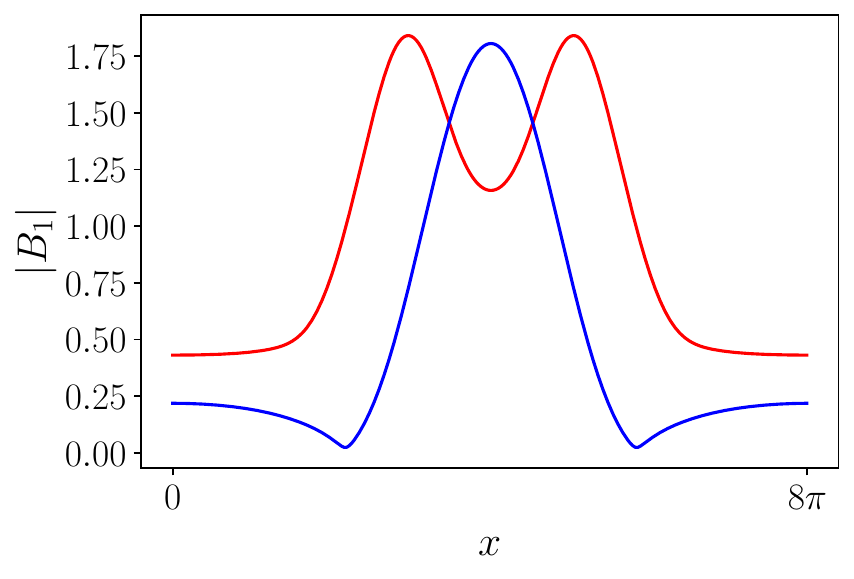}
    \put(0,65){(a)}
    \end{overpic}~~
   \begin{overpic}[width=0.45\linewidth]{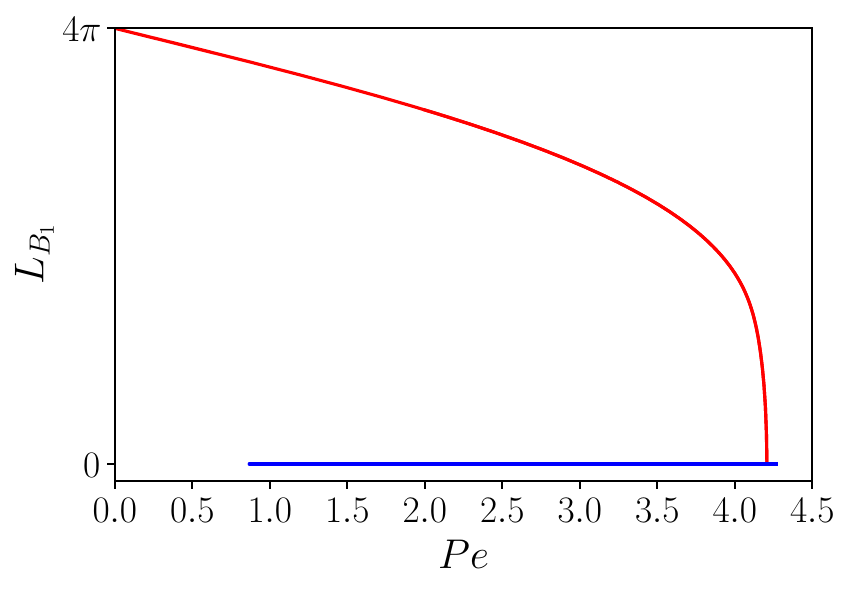}
    \put(-3,65){(b)}
    \end{overpic}
  \caption{\label{fig:delta} (a) The value of $|B_1|(x)$ for the UB solution, $\bm{X}_{\rm UB}$, (red, double peak) and 
  for the LB solution, $\bm{X}_{\rm LB}$, (blue, single peak). (b) Distance between the two peaks in $|B_1|$ of the UB solution as a function of $Pe$. }
\end{figure}

We also show the complex amplitudes in the vorticity and density deviation fields in Fig.\ref{fig:LB}(c,d). Notably, the $B_1$ amplitude does not possess the peaks around the kinks of the localized field [Fig.\ref{fig:LB}(d)]. This difference are clearly shown in the plot of $|B_1|$ in Fig.\ref{fig:delta}(a), as seen in the double-peak structure for the UB solution (red) and single-peak structure for the LB solution (blue).

To further quantify these differences, we introduce the distance between the two peaks in $|B_1|$ as $L_{B_1}$. Hence, $L_{B_1}=0$ when two peaks merge, and we found $L_{B_1}=0$ for all LB solutions. The value of $L_{B_1}$ is then calculated for the UB solution by estimating the peak position using the three-point Lagrange interpolation method, the points of the grid being equally separated into 256 parts. As shown in Fig. \ref{fig:delta}, the value of $L_{B_1}$ for $\bm{X}_{\rm UB}$ decreases monotonically to reach zero around $Pe\approx Pe_{\rm SN}$.

\subsection{Edge state and basin boundary}

In the parameter regime discussed earlier in this section, we have observed the bistable nature of the system. We then further examine the global stricture of the dynamical system.

\begin{figure}[tb!]
  \begin{overpic}[width=0.45\linewidth]{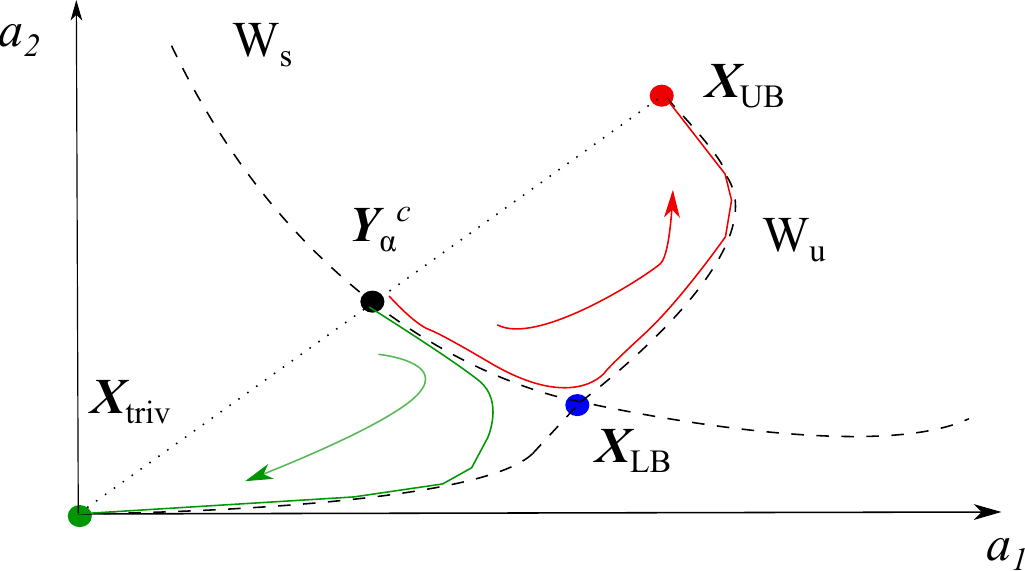}
    \put(0,63){(a)}
    \end{overpic}
   \begin{overpic}[width=0.45\linewidth]{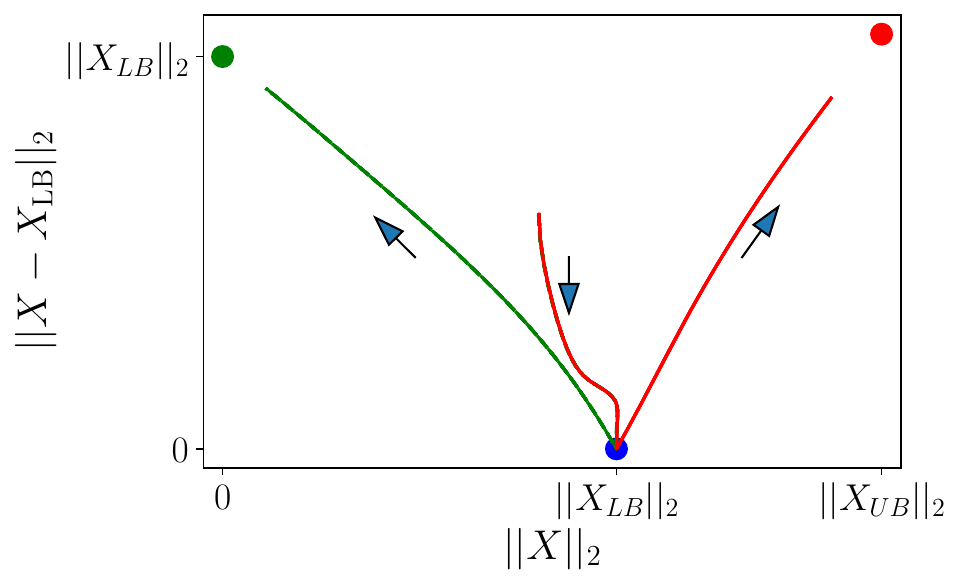}
    \put(0,63){(b)}
    \end{overpic}
  \caption{\label{fig:bisec2} (a) Schematic of the global bistable structure of the dynamical system projected onto an two-dimensional state space.
  The trivial solution and the UB solution attract fixed points, and the LB solution behaves as an edge state; the unstable manifold ($W_u$) connects to each stable fixed point, and the stable manifold ($W_s$) forms a basin boundary between the trivial solution and the UB solution. We numerically detect the basin boundary by finding an interior point in the segment between the trivial and UB solutions, $\bm{Y}_\alpha^c$, from which the orbit approaches the UB solution. The axis labels $a_1$ and $a_2$ represent two representative bases of the state space. 
  (b) The computed orbits of the dynamical system projected on to the two-dimensional state plane. 
  The trivial solution and the UB solution are depicted by green and red circles, 
  and the LB solution is marked by a blue circle. 
  The arrows indicate time evolution, and the colors correspond to the orbits in (a).  
  }
\end{figure}

By analyzing the time evolution around the unstable UB solution, we numerically observed some orbits stay longer at the neighborhood. 
We then chased the generated orbits and using the bisection method (details are explained later), we have numerically detected the basin boundary by which the final asymptotic state is separated. In Fig.\ref{fig:bisec2}(a), we show schematic of the basin boundary and the global structure of the steady solutions. While the orbits in the state space are attracted to the trivial solution or the UB solution, the LB solution behaves as an edge state \cite{itano_dynamics_2001,Khapko_2013}, which separates the attracting state of orbits. As schematically shown in the figure, the unstable manifold ($W_u$) connects to each stable fixed point, and the stable manifold ($W_s$) forms a basin boundary between the trivial solution and the UB solution.

\begin{figure}[tb!]
  \begin{overpic}[width=0.45\linewidth]{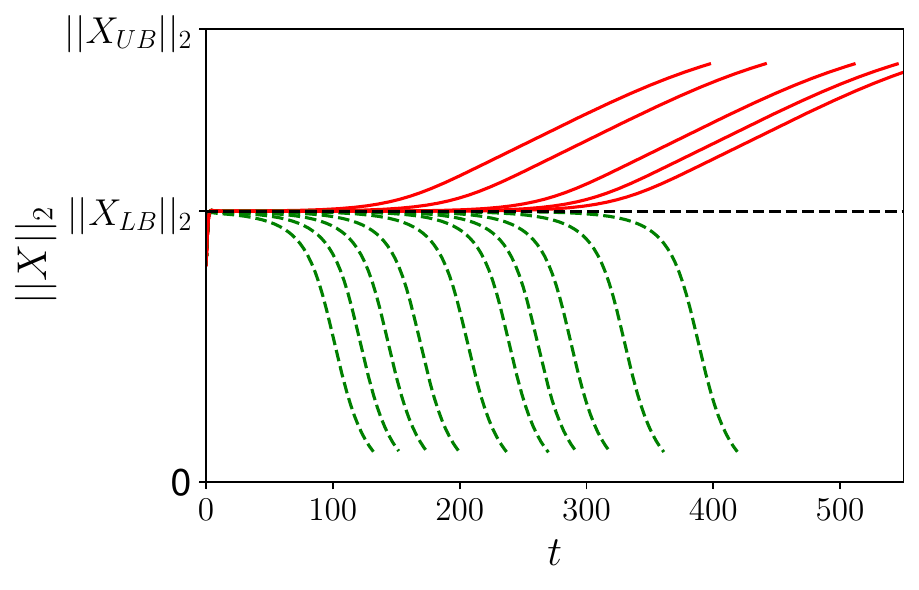}
    \put(0,65){(a)}
    \end{overpic}~~
   \begin{overpic}[width=0.45\linewidth]{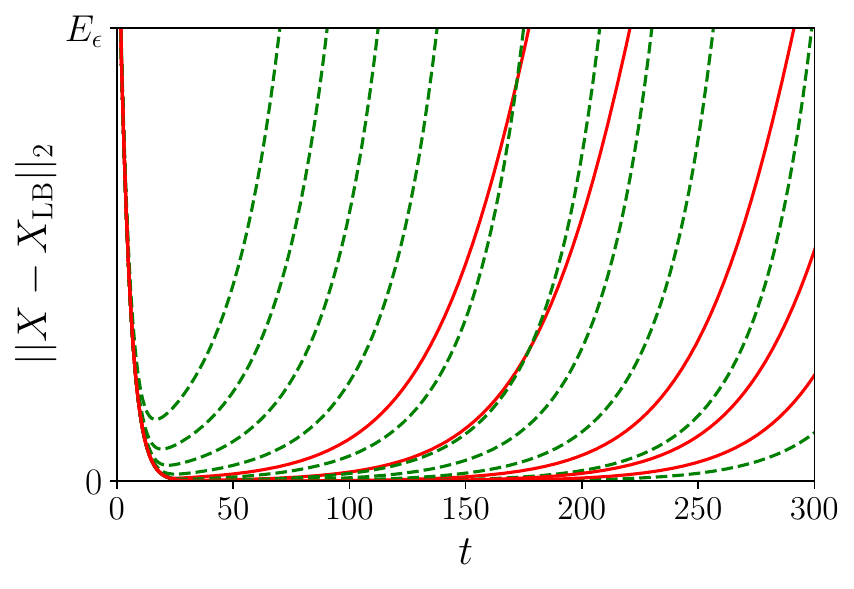}
    \put(-2,65){(b)}
    \end{overpic}
  \caption{\label{fig:bisec} The basin boundary computed by the bisection method. (a) A set of trajectories that finally reach the trivial solution (green) and the UB solution (red).
  The basin boundary is obtained as the broken line. 
  (b) The relative distance to the LB solution for the same trajectories as in (a). The states initially located near the basin boundary stay longer at the neighborhood of the LB solution.}
\end{figure}

Indeed, we have numerically obtained these structures and the results are shown in Fig.\ref{fig:bisec2}(b), where the computed trajectories are projected onto a two-dimensional state space that is spanned by the norms
$(||X||_2,||X-X_{\rm LB}||_2)$. 
The two stable states are shown in the colored circles: the trivial solution marked by a green circle 
at the upper-left corner 
and the UB solution marked by a red circle at the right-upper corner.
The two representative  trajectories are shown in different colors,
and the green and red curves show orbits reaching the trivial and UB solutions, respectively, corresponding the schematic trajectories in Fig.\ref{fig:bisec2}(a).
The dotted lines represent the basin boundary that is numerically obtained by the bisection method.

We now briefly describe the bisection method that is used to detect the basin boundary. We first consider a state as a linear interpolation of the two steady solution as $\bm{Y}_{\alpha}=\alpha \bm{X}_{\rm UB}(\alpha \in [0,1])$.
Our aim is then to find $\alpha^c \in [0,1]$, around which the eventual states are separated [Fig.\ref{fig:bisec2}(a)]. The numerical results show that
the orbits stay at the neighborhood of the LB solution, $\bm{X}_{LB}$, and depart exponentially in time, as shown in the schematic orbits [Fig.\ref{fig:bisec}(a)] and in the projection [Fig.\ref{fig:bisec}(b)].
Further details are shown in Fig.\ref{fig:bisec}, where we show several orbits near the basin boundary. 
In Fig.\ref{fig:bisec}(a) we present a set of orbits that finally reach the trivial solution (green) 
and the UB solution (red), 
and the basin boundary is then obtained as the broken line.
The plots in Fig.\ref{fig:bisec}(b) show the relative distance to the LB solution for the same trajectories and we find that the states initially located nearer to the basin boundary stay longer at the neighborhood of the LB solution.

\section{Discussion and conclusions \label{sec:conclusion}}

In this study, motivated by localized structure and bistability of bioconvection patterns, we investigated nonlinear fluid-density couplings in a model convection system. To focus on the bulk properties, we introduced an equilibrium density profile as an independent model parameter, and we extended a well-studied bioconvection model to fit with a general boundary condition. Hence, our system contains the self-propulsion of the particle as an independent nondimensional parameter $Pe$. 

The key nonlinear coupling lies between the incompressible fluid motion and particle density fields. Since the dynamics are only dominated by the largest-scale modes in the vertical direction, we were able to introduce a system that truncates smaller vertical structure without altering its topological features of the dynamical system, which are also confirmed by direct computations of the full model.

By examining the reduced system, we analytically derived the neutrally stable curve of the linear stability [Eq.\eqref{eq:N}] and found that the system stabilizes to the base flow at a large $Pe$. We then performed bifurcation analyzes and found the bistable structure of the horizontally localized convection pattern, associated with a downward fluid velocity
[Fig.\ref{fig:bifurcation_sc25}]. 
These findings, localization and bistability,
are consistent with existing experimental studies of bioconvection \cite{suematsu_2011,shoji_2014,yamashita_2023}, highlighting that these structures can emerge only by particle-biased self-propulsion and do not rely on detailed biological reactions
such as cell-wall and cell-cell interactions.
In the experiment by Shoji et al. \cite{shoji_2014}, it is reported that the localized convection moves slowly in the horizontal direction. In our study, however, we have found by a full model simulation that all the obtained localized pattern does not move and hence we only analyzed non-travelling solutions by imposing the reflection symmetry in the horizontal direction.

The neutrally stable curve obtained, Eq.\eqref{eq:N}, indicates that the linear instability is also induced by an increase of $Ra$, since the critical P\'eclet number, $Pe^{\rm c}$ [Eq.\eqref{eq:Pec}], increases as $Ra$. The bifurcation diagram therefore indicates that the bistability structure emerges above the critical $Ra$.  
Since the Rayleigh number, $Ra$, represents a nondimensional equilibrium particle density, this result physically means that the stable convection pattern should be realized above a critical suspension density.

Another notable feature of the bifurcation diagram is the disappearance of a steady convection state at a higher P\'eclet number ($Pe>Pe_{\rm SN}$). This result apparently seems inconsistent with existing experimental and numerical observations where biological tactic behavior induces the bioconvection instability\cite{taheri_bioconvection_2007,kasyap_2012,bees_2020},
although these differences results from the different model formulations. In our extended model, we considered the equilibrium density profile, $m_0$, as an independent parameter to capture the nonlinear effects driven by the particle self-propulsion. 
However, in a typical experimental setup, due to the density condition of no flux at the material boundary, the equilibrium density profile, or the Rayleigh number $Ra$, is no longer an independent parameter, meaning that $m_0$ is a function of $Pe$. 
More precisely, it should follow $dm_0/dy \propto Pe\exp(Pe y)/[\exp(Pe L_y)-1]$, where $L_y=2\pi$ is the vertical size of the fluid region.
This nontrivial dependence of the particle speed on the system parameters usually makes it difficult to interpret numerical and experimental results and thereby to understand the underlying mechanism. In contrast, our extended formulation enabled clearer analysis and interpretation of the nonlinear coupling in a bulk.

To further characterize the global structure of the bistable dynamical system, we have numerically detected the stable and unstable manifolds of the unstable steady state and found that this steady solution behaves as an edge state in a sense that its stable manifold globally separates the state space into two basins of attraction, corresponding to the two stable steady solutions. The existence of the edge state characterizes the bi-stability of the bioconvection. Further, the global connection between the steady states enables one to visualize the time-evolving process of the creation and annihilation of the convection pattern. This also quantifies the minimum disturbance to generate the localized convection pattern from the quiescent state. 

\begin{figure}[tb!]
  \begin{overpic}[width=0.45\linewidth]{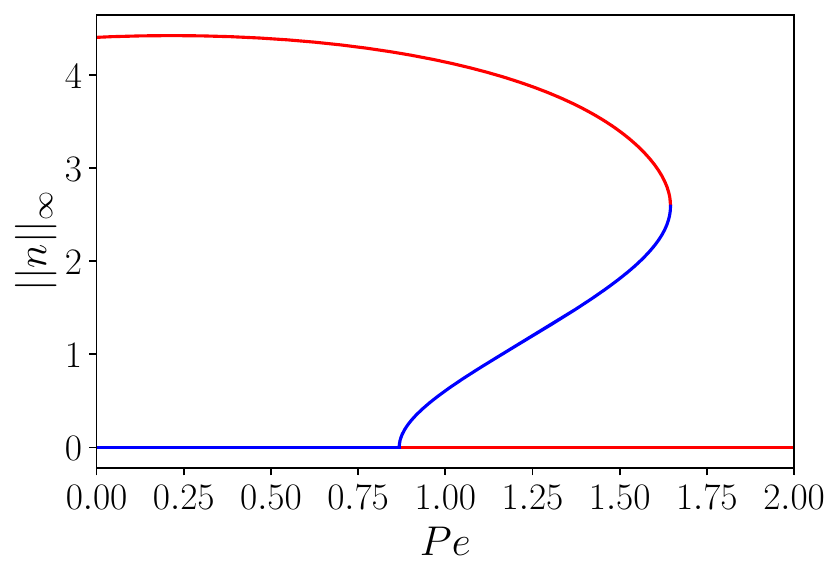}
    \put(0,65){(a)}
    \end{overpic}
   \begin{overpic}[width=0.45\linewidth]{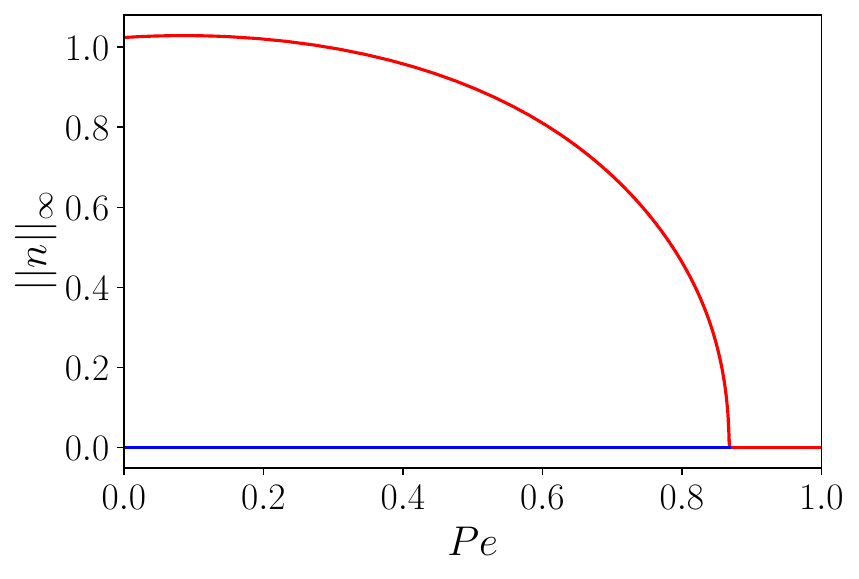}
    \put(0,65){(b)}
    \end{overpic}
  \caption{\label{fig:bifurcation_scs}  Bifurcation diagram of the steady solution at smaller $Pr$ values. (a) $Pr=1$ and (b) $0.4$. We used the same value of $Ra$ as in Fig.\ref{fig:bifurcation_sc25}(a).}
\end{figure}

The results shown in this paper focused on a biologically reasonable parameter value of $Pr$, which physically represents an inverse of nondimensional particle diffusion. In Fig. \ref{fig:bifurcation_scs}, we explored the bifurcation diagram at smaller values of $Pr$. 
From the neutrally stable curve, we found that the value of $Pr$ does not alter the linear stability. The nonlinear stability, however, are altered by the change of $Pr$. As shown in Fig. \ref{fig:bifurcation_scs}(a), when we decrease the Prandtl number from $Pr=2.5$ to $Pr=1$, the saddle-node bifurcation occurs at a lower value of $Pe$. A further decrease of $Pr$ to $Pr=0.4$ finally loses the saddle-node bifurcation and the bistable structure [Fig. \ref{fig:bifurcation_scs}(b)]. There only exists one stable steady solution for all the $Pe$ values. Biological data from Paramecium \cite{taheri_bioconvection_2007} report a large variation of $Pr$, suggesting that this transition could be experimentally observed.

Nevertheless, our study relies on the two-dimensional assumption of the fluid motion and
it is not clear whether the same scenario follows for the fully three-dimensional system. 
For example, the two-dimensional Navier-Stokes system under a large-scale forcing is known to exhibit unimodal flow pattern that covers the whole fluid region even in the turbulent regime \cite{Kraichnan_1959,Gallet_2013,Kim_2015,sasaki_2020}. 
In our model convection system, we assumed an equilibrium density profile that vertically varies over the system size, and
a similar large-scale flow pattern was realized through the bulk nonlinear coupling
Hence, the emerging flow patterns in the three dimensions and with a small-scale forcing may have complex structures, and this problem is left for future work.

In the current system, we numerically found that there cannot exist stably a convection cell structure with finite wavelength and that the long-wavelength flow structure dominates. Further, we numerically confirmed that no homoclinic snaking bifurcations occur in our system, although this bifurcation is often observed in other convection and Swift-Hohenberg systems \cite{burke2007,Watanabe_2012}. 
Our convection solution bifurcates from a horizontally homogeneous trivial solution, and this is consistent with non-snaking conditions of
Beaume et al. \cite{Beaume2013}, where
they examined convection solutions with kink- and anti-kink structure as in our system.

In conclusion, with an extended model of bioconvection, we have found that the particle self-propulsion can induce a localized, bi-directional convection pattern, regardless of boundary effects and biological detailed reactions, highlighting the importance of fluid-density nonlinear coupling in bioconvection systems. Further, we have successfully characterized the global structure of the bistable dynamical system as the emergence of an edge state. The current findings deepen our understanding of the role of hydrodynamics and particle activity in bioconvection. More generally, our theoretical methodology based on the dynamical systems theory are useful in understanding complex flow patterns in nature.

\section*{Acknowledgements}
The authors acknowledge Dr. Hiroshi Yamashita and Prof. Makoto Iima for a fruitful discussion.
K.I. acknowledges the Japan Society for the Promotion of Science (JSPS) KAKENHI for Transformative Research Areas A (Grant No. 21H05309) and the Japan Science and Technology Agency (JST), FOREST (Grant No. JPMJFR212N). Y.H. and K.I. were supported in part by the Research Institute for Mathematical Sciences, an International Joint Usage/Research Center located at Kyoto University.

 \bibliography{bib}

\end{document}